\documentclass[a4paper,12pt]{article}
\usepackage{amssymb,amsmath}
\usepackage{graphics}
\def\beq{\begin{equation}}
\def\eeq{\end{equation}}
\def\bea{\begin{eqnarray}}
\def\eea{\end{eqnarray}}
\def\nn{\nonumber}
\makeatletter
  \def\@cite#1#2{${\mbox{#1\if@tempswa , #2\fi}}$}
\makeatother

\makeatletter
  \def\@biblabel#1{$^{\mbox{#1}}$}
\makeatother
\begin{document}
%
%
%
%
\thispagestyle{empty}
\vspace*{3cm}
\begin{center}
{\LARGE\sf A Fractional entropy in Fractal phase space: properties and characterization} \\

\bigskip\bigskip
R. Chandrashekar${}^{1}$, C. Ravikumar${}^{2}$ and J. Segar${}^{3}$

\bigskip
\textit{
${}^{1}$The Institute of Mathematical Sciences, \\
C.I.T Campus, Taramani, \\
Chennai 600 113, India \\}
\bigskip
\textit{
${}^{2}$ Department of Theoretical Physics, \\
University of Madras, \\
Maraimalai Campus, Guindy, \\
Chennai 600 025, India \\}
\bigskip
\textit{
${}^{3}$Ramakrishna Mission Vivekananda College\\
Mylapore\\
Chennai 600 004, India.
}
\end{center}

\vfill
\begin{abstract}
A two parameter generalization of Boltzmann-Gibbs-Shannon entropy based on natural logarithm is introduced.
The generalization of the Shannon-Kinchinn axioms corresponding to the two parameter entropy is proposed and
verified. We present the relative entropy, Jensen-Shannon divergence measure and check their properties. 
The Fisher information measure, relative Fisher information and the Jensen-Fisher information corresponding 
to this entropy are also derived. The canonical distribution maximizing this entropy is derived and is found
to be in terms of the Lambert's $W$ function.  Also the Lesche stability and the thermodynamic stability 
conditions are verified. Finally we propose a generalization of a complexity measure and apply it to a two level 
system and a system obeying exponential distribution.  The results are compared with the corresponding ones
obtained using a similar measure based on the Shannon entropy.    
\end{abstract}

PACS Number(s):  
\newpage
\setcounter{page}{1}
%
%
%
\setcounter{equation}{0}
\section{Introduction}
\label{Intro}
Entropy is a very important quantity and plays a key role in many aspects of statistical mechanics and information
theory.  The most widely used form of entropy was given by Boltzmann and Gibbs from the statistical mechanics point 
of view and by Shannon from an information theory point of view. Later certain other generalized measures of entropy
like the R\'enyi entropy [\cite{AR61}] and the Sharma-Mittal-Taneja entropy [\cite{DM75},\cite{BS75}] were introduced 
and their information theoretic aspects were investigated. Recently in [\cite{CT88}] a new expression for the entropy 
was proposed as a generalization of the Boltzmann-Gibbs entropy and the necessary properties like concavity, Lesche
stability and thermodynamic stability were verified. This entropy has been applied to a wide variety of physical 
systems, in particular to long-range interacting systems [\cite{BC02},\cite{AP04}] and nonmarkovian systems [\cite{AM11}].

Most of the generalized entropies introduced so far were constructed using a deformed logarithm. But two generalized 
entropies one called the fractal entropy and the other known as fractional entropy were proposed using the natural 
logarithm.  The fractal entropy which was introduced in [\cite{QAW03}] attempts to describe complex systems which 
exhibit fractal or chaotic phase space.  Similarly the fractional entropy was put forward in [\cite{MR98}] and later
applied to study anomalous diffusion [\cite{MR09}].  Merging these two entropies in our present work we propose 
a fractional entropy in a fractal phase space.  Thus there are two parameters one parameter characterizing the 
fractional nature of the entropy and the other parameter describing the fractal dimension of the phase space. 
Thus the functional form of the entropy depends on the natural logarithm.  We give the generalized 
Shannon-Kinchinn axioms corresponding to this two parameter entropy and prove that they uniquely characterize
our entropy.  The two parameter relative entropy and the Jensen Shannon divergence measure are also generated. 
The generalized Fisher information is derived from the relative entropy.  Relative Fisher information measure 
and its associated Jensen-Fisher information measure corresponding to this entropy are also proposed.  The 
thermodynamic properties like the Lesche stability and the thermodynamic stability are also verified.  We 
notice that the probability distribution which maximizes this entropy is expressible in terms of the Lambert's
$W$-function.  Finally we set up a two parameter generalization of the well known complexity measure 
LMC (L\'{o}pez-Ruiz, Mancini and Calbet) complexity measure [\cite{RL95}] and apply it to a two level system 
and an exponential distribution.

After the introduction in Section I, we introduce our new two parameter entropy in Section II and 
investigate its properties.  In Section III the relative entropy and the Jensen-Shannon divergence measure 
corresponding to this two parameter entropy is presented and its properties are studied. Using the relative
entropy the Fisher information measure, the relative Fisher information and the Jensen-Fisher information are also obtained. 
The thermodynamic properties are analyzed in Section IV. In section V we present a two 
parameter generalization of the LMC complexity measure and analyze the complexity of a two level system and 
a system with continuous probability distribution. We conclude in Section VI.  

%
%
%
\setcounter{equation}{0}
\section{Generalized entropy and its axiomatic characterization}
\label{axioms}
The Boltzmann-Gibbs-Shannon entropy which is an expectation value of $\ln (1/p_{i})$ is 
generally expressed as 
\beq
S_{BG} = \langle \ln (1/p_{i}) \rangle   \equiv  k \ \sum_{i} p_{i}  (- \ln p_{i}), 
\label{entr_BGS}
\eeq
where $p_{i}$ represents the probability and $k$ is a constant. A new form of entropy based on the 
natural logarithm was proposed in [\cite{QAW03}] considering $p_{i}^{q}$ as the effective probability i.e.,
$\sum_{i} p_{i}^{q} = 1$ to take into account incomplete information. The entropy thus defined 
\beq
S = \langle \ln (1/p_{i}) \rangle_{q} \equiv \sum_{i} p_{i}^{q} (- \ln p_{i}),
\label{entr_fractal}
\eeq
makes use of the $q$-expectation given below:
\beq
\langle O \rangle_{q} = \sum_{i} p_{i}^{q} O.
\label{q_expec}
\eeq
The $q$-expectation value (\ref{q_expec}) characterizes incomplete normalization [\cite{QAW03}] which is known to 
occur in complex systems. Later to account for the mixing which occurs due to interactions between the various 
states of the system, the same form of the entropy but with the regular conditions on the probabilities i.e., 
$\sum_{i} p_{i} = 1$  was discussed in [\cite{FS04},\cite{FS09}].

The Boltzmann-Gibbs-Shannon entropy can also be defined through the equation
\beq
S_{BG} = - \frac{{\rm d} }{{\rm d}x}  \sum_{i} p_{i}^{x}\bigg{|}_{x=1}.
\eeq
Replacing the ordinary derivative by the Weyl fractional derivative a new entropy was obtained by Ubriaco
in [\cite{MR98}].  The functional form of the entropy which is an expectation value of $(\ln (1/p_{i}))^{q}$ reads:  
\beq
S = \sum_{i} p_{i} (- \ln p_{i})^{q} \equiv \langle (\ln (1/p_{i}))^{q} \rangle.
\label{entr_fractional}
\eeq
A salient feature of the fractal entropy (\ref{entr_fractal}) and the fractional 
entropy (\ref{entr_fractional}) is that they are functions of the ordinary logarithm unlike the
other generalized entropies [\cite{CT88},\cite{GK01},\cite{AL07}] which are defined through the use of deformed logarithms.  

Inspired by fractal entropy (\ref{entr_fractal}) and the fractional entropy (\ref{entr_fractional}) we 
propose a two parameter generalization of the Boltzmann-Gibbs-Shannon entropy
\beq
S_{q,q^{\prime}} (p_{i})= k \langle (\ln (1/p_{i}))^{q^{\prime}} \rangle_{q} 
		 \equiv k \sum_{i=1}^{W} p_{i}^{q}\left( - \ln p_{i}\right)^{q^{\prime}} 
                  \equiv k\sum_{i=1}^{W} p_{i}^{q}\left(\ln\frac{1}{p_{i}}\right)^{q^{\prime}},
\label{2p_entr}
\eeq
where $k$ is a generalization of the Boltzmann constant and $q$ and $q^{\prime}$ are the 
parameters which are used to generalize the BGS entropy. The entropy (\ref{2p_entr}) can be 
considered as a fractional entropy in a fractal phase space in which the parameter $q$ comes from the 
the fractal nature and the parameter $q^{\prime}$ is from the fractional aspect. 

In the $q \rightarrow 1$ limit the entropy (\ref{2p_entr}) reduces to the fractional entropy (\ref{entr_fractional}).
Similarly we recover the fractal entropy (\ref{entr_fractal}) in the limit $q^{\prime} \rightarrow 1$ and the 
BGS entropy when both the parameters attain the value of unity. A very interesting limiting case of the 
(\ref{2p_entr}) occurs when we set $q=q^{\prime}$ 
\beq
S_{q} (p_{i})= k \, \sum_{i=1}^{W}  p_{i}^{q}\left(-\ln p_{i}\right)^{q} 
              \equiv k \, \sum_{i=1}^{W} p_{i}^{q}\left(\ln\frac{1}{p_{i}}\right)^{q}
             \equiv k \, \sum_{i=1}^{W} (s_{i}^{B})^{q},
\label{Ub_Fr_entr}
\eeq
where $s_{i}^{B}$ is the single particle Boltzmann entropy. The one parameter entropy (\ref{Ub_Fr_entr}) is the 
sum of biased single particle Boltzmann entropy.  At this juncture we would like to make a remark about the 
Boltzmann entropy and the Gibbs entropy. An explanation in [\cite{EJ65}] states that the Boltzmann entropy is the
$N$ sum of the entropy calculated from the one particle distribution, whereas the Gibbs entropy is computed 
directly from the $N$-particle distribution.  This implies that the Boltzmann entropy and the Gibbs entropy are
the same only when the systems are noninteracting. Looking into equation (\ref{Ub_Fr_entr}) from this point of view we
realize that this entropy can be understood in a similar setting, i.e., the one parameter entropies are biased
by a parameter $q$ and this bias may be due to the presence of interactions.  Such a behaviour strongly resembles
the characteristics of complex systems in which the behaviour of the total system is different from the single 
particle system due to presence of interactions.  So we assume that the entropy (\ref{Ub_Fr_entr}) described above 
may be a strong candidate in describing complex systems.  

Below we present the two parameter generalization of the Shannon-Kinchinn axioms. 
Let $\Delta_{n}$ be an $n$-dimensional simplex as defined below
\beq
\Delta_{n} = \bigg\{(p_{1},\ldots, p_{n}) \bigg{|} \, p_{i} \geq 0, \sum_{i=1}^{n} p_{i} = 1 \bigg\},
\eeq
in which the two parameter entropy (\ref{2p_entr}) satisfies the following axioms. \\
{\it (i) Continuity:}  The entropy $S_{q,q^{\prime}}$ is continuous in $\Delta_{n}$.  \\
{\it (ii) Maximality:} For any $n \in N$ and any $(p_{1},\ldots,p_{n}) \in \Delta_{n}$
\beq
S_{q,q^{\prime}}(p_{1},\ldots , p_{n}) \leq S_{q,q^{\prime}} \left(\frac{1}{n},\ldots,\frac{1}{n}\right).
\eeq
{\it (iii) Expansibility:}
\beq
 S_{q,q^\prime}(p_{1},....,p_{n},0) = S_{q,q^\prime}(p_{1},....,p_{n}).
\eeq
{\it (iv) Generalized Shannon additivity:} 
\bea
S_{q,q^\prime}(p_{ij}) &=& \frac{1}{2}\Big(\sum_{i} s_{q,q^\prime}(p_{i}) \sum_{j}(p(j|i))^q 
                            + \sum_{i} p_{i}^q \sum_{j}s_{q,q^\prime}(p(j|i))   
\label{Gen_shn_2p}\\
                        & & + \sum_{k=1} \binom{q^{\prime}}{k} \Big(\sum_{i} s_{q,k}(p_{i}) 
                            \sum_{j}s_{q,q^\prime-k} (p(j|i)) +\sum_{i} s_{q,q^\prime-k}(p_{i}) 
                            \sum_{j}s_{q,k} (p(j|i)) \Big)\Big)\nn
\eea
where 
\beq
s_{q,q^{\prime}}(p_{i})= p_{i}^q (-\ln p_{i})^{q^{\prime}}  \qquad \qquad
{\displaystyle{\sum_{i}}} s_{q,q^{\prime}}(p_{i})= S(p_{i}).
\eeq
The factor  $p(j|i) = p_{ij}/p_{i}$ appearing in (\ref{Gen_shn_2p}) is the conditional probability 
i.e., probability of occurence of a $j^{th}$ event when a particular $i^{th}$ event has occured.
\begin{figure}[h!]
\begin{center}
\resizebox{75mm}{!}{\includegraphics{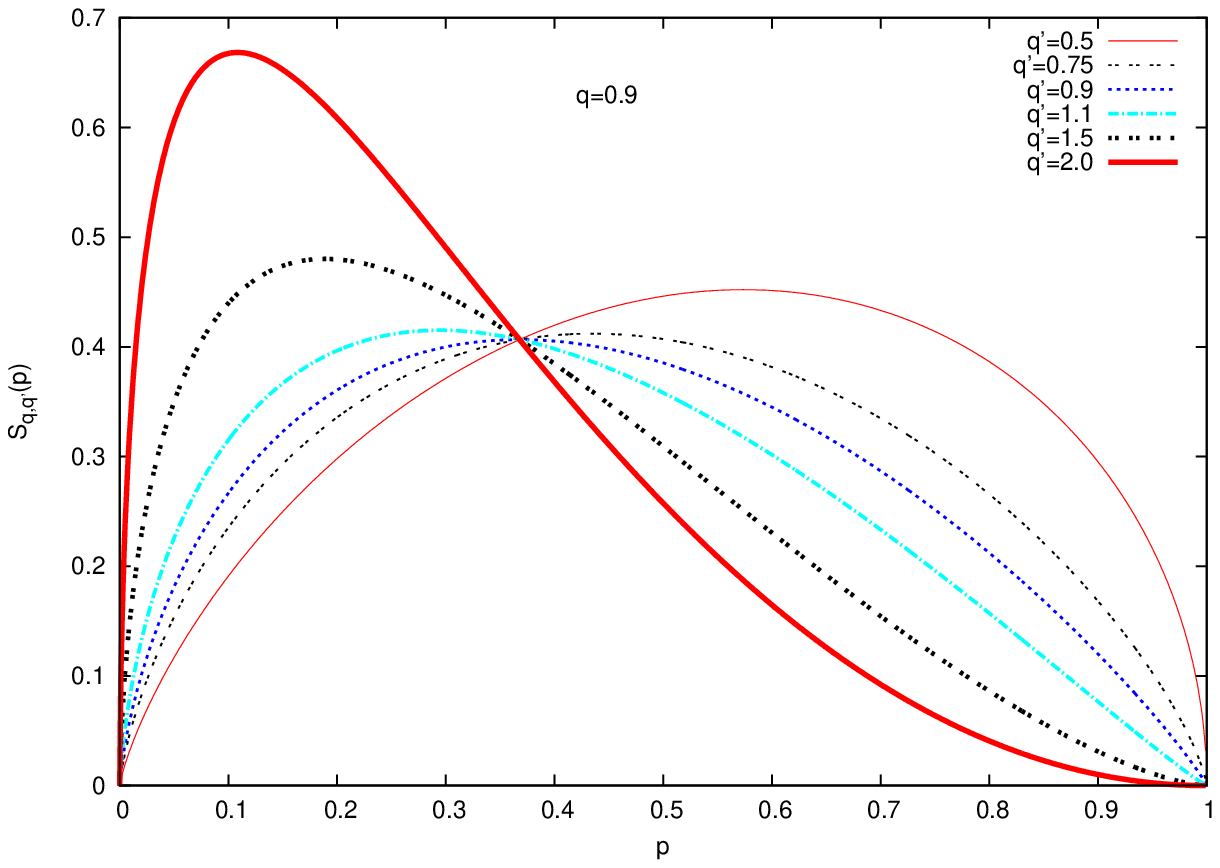}}
\resizebox{75mm}{!}{\includegraphics{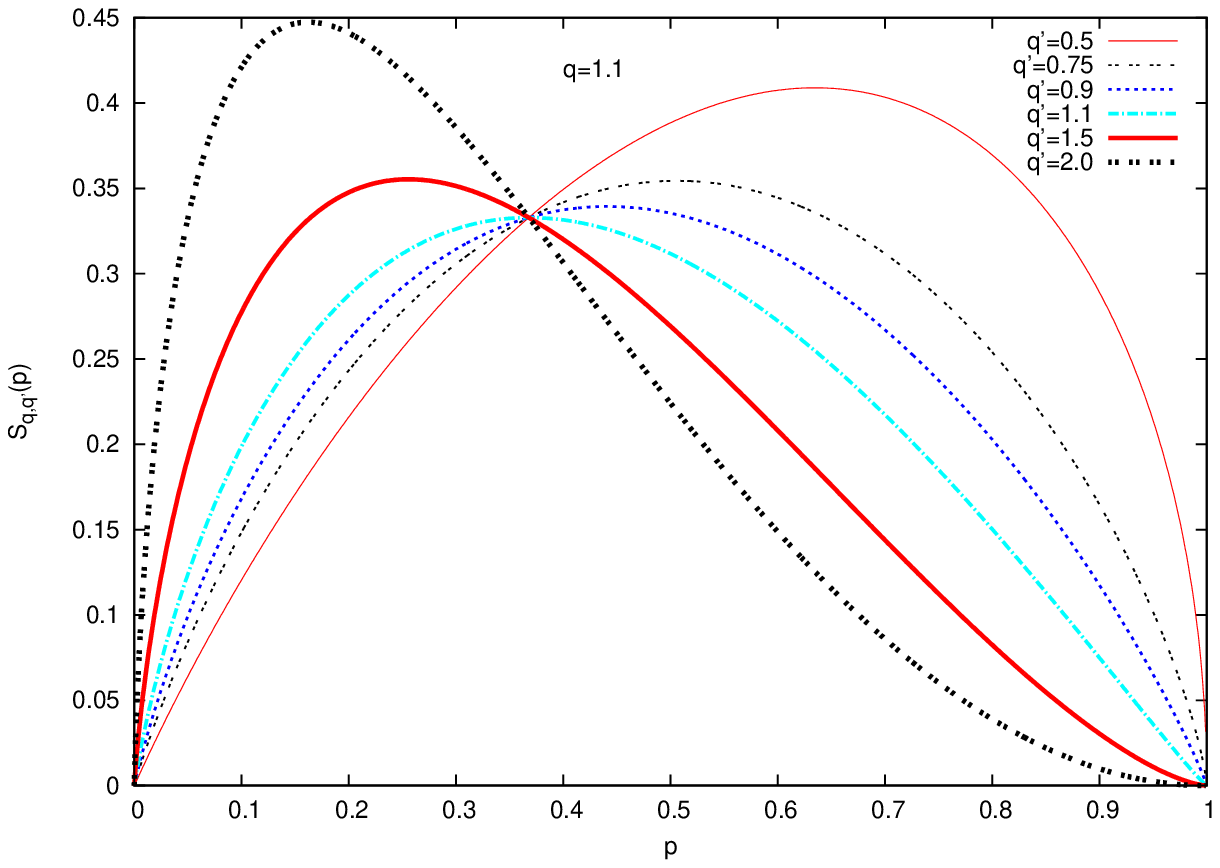}}
\resizebox{75mm}{!}{\includegraphics{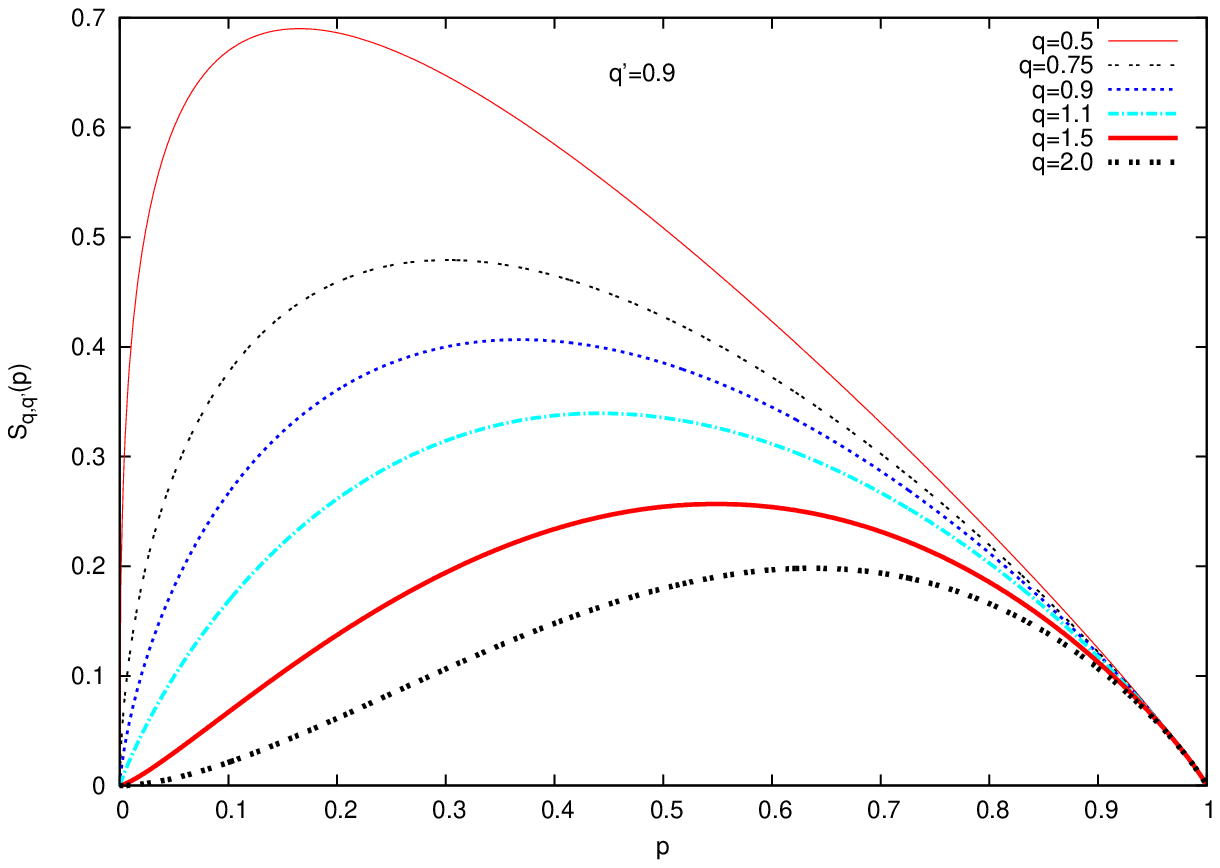}}
\resizebox{75mm}{!}{\includegraphics{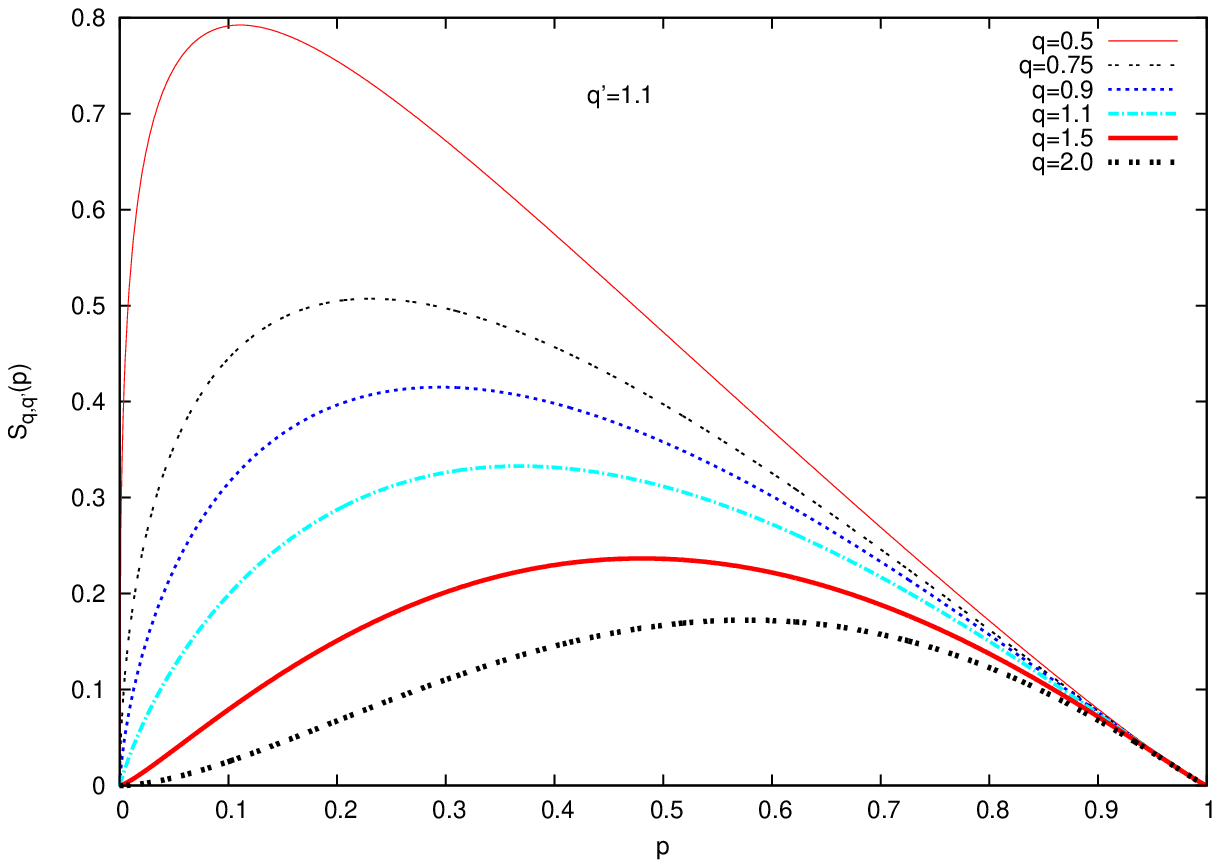}}
\end{center}
\caption{In the following graphs we show that the entropic function defined in (\ref{2p_entr}) is concave for various
values of $q$ and $q^{\prime}$. The first two graphs in the top row are plotted for various values of $q^{\prime}$ keeping 
the value of $q$ fixed at $0.9$ and $1.1$ respectively. The next set of graphs are plotted for various values of $q$ keeping
the value of $q^{\prime}$ fixed at $0.9$ and $1.1$.}
\end{figure} 
\subsection{Concavity:}
The two parameter entropic functional (\ref{2p_entr}) has an extremum at $p_{i} = \exp(-q/q^{\prime})$ and the 
second derivative w.r.t $p_{i}$ is
\beq
\frac{\partial^{2} S}{\partial p_{i}^2}\bigg{|}_{p_{i}=   e^{-q/q^{\prime}}} 
             = - q^{\prime} \, \exp\left(\frac{q^{\prime} (2 - q^{\prime})}{q} \right) \, 
                 \left(\frac{q^{\prime}}{q}\right)^{q^{\prime}-2}.
\label{2p_2deriv_p}
\eeq
From (\ref{2p_2deriv_p}) it can be observed that the second derivative of the entropy is uniformly $-$ve 
in the region $(q,q^{\prime}) > 0$ implying that the entropic functional is uniformly concave for 
$q,q^{\prime} \in \mathbb{R}^{+}$. The various limiting cases arising out of the two parameter entropy 
{\it viz} the fractal entropy (\ref{entr_fractal}), the fractional entropy (\ref{entr_fractional}) and 
the one parameter entropy (\ref{Ub_Fr_entr}) are also concave only when the deformation parameters are 
greater than zero.  We illustrate the concavity of the entropic function through a set of plots 
shown above. 

\subsection{Generalised form of Shannon additivity}
Let $p_{ij}$ be the probability of occurence of a joint event, in which $p_{i}$ and $p_{j}$ are probability
of occurence of the individual events. The two parameter entropy corresponding to the joint probability
$p_{ij}$ can be written as
\beq
S_{q,q^\prime}(p_{ij}) = \sum_{i,j} p_{ij}^{q}\left(-\ln p_{ij}\right)^{q^{\prime}} 
                        = \sum_{i,j} \big(p(j|i)\big)^{q} p_{i}^{q} 
                          \bigg(-\ln \left(p_{i} \ p(j|i) \right)\bigg)^{q^{\prime}}.
\label{entr_jp}
\eeq
Expanding the logarithm in the above equation using the binomial theorem and isolating the $k=0$ term we arrive at
\beq
S_{q,q^\prime}(p_{ij}) = \sum_{i} s_{q,q^\prime}(p_{i}) \sum_{j} \big(p(j|i)\big)^{q}  
                         + \sum_{k=1}  \binom{q^{\prime}}{k}  \sum_{i} s_{q,q^\prime-k}(p_{i}) 
                           \sum_{j} s_{q,k}\big(p(j|i)\big),
\label{bino_exp_1}                      
\eeq
where $\binom{q^{\prime}}{k}$ denotes the binomial coefficients.  
Since the system is symmetric in $p_{i}$ and $p(j|i)$, the binomial expansion of eqn (\ref{entr_jp}) 
can be written in an equivalent form as
\beq
S_{q,q^\prime}(p_{ij})= \sum_{i} p_{i}^q \sum_{j} s_{q,q^\prime}(p(j|i)) 
                        + \sum_{k=1} \binom{q^{\prime}}{k}  \sum_{i} s_{q,k}(p_{i}) 
                        \sum_{j} s_{q,q^\prime-k}\big(p(j|i)\big).
\label{bino_exp_2}
\eeq
Adding (\ref{bino_exp_1}) and (\ref{bino_exp_2}) the modified form of Shannon additivity is obtained 
\bea
S_{q,q^\prime}(p_{ij}) &=& \frac{1}{2}\Big(\sum_{i} s_{q,q^\prime}(p_{i}) \sum_{j}(p(j|i))^q 
                            + \sum_{i} p_{i}^q \sum_{j}s_{q,q^\prime}(p(j|i))   
\label{Gen_shn_addt}\\
                        & & + \sum_{k=1} \binom{q^{\prime}}{k} \Big(\sum_{i} s_{q,k}(p_{i}) 
                            \sum_{j}s_{q,q^\prime-k} (p(j|i)) +\sum_{i} s_{q,q^\prime-k}(p_{i}) 
                            \sum_{j}s_{q,k} (p(j|i)) \Big)\Big)\nn
\eea
When the two events are independent i.e., the joint probability obeys the relation \\
$p_{ij} = p_{i} p_{j}$ eqn (\ref{Gen_shn_addt}) simplifies into 
\bea
S_{q,q\prime}(p_{ij}) &=& \frac{1}{2} \Biggl(S_{q,q^{\prime}}(p_{i}) + S_{q,q^{\prime}}(p_{j}) 
                       - M_{j}(q) \, S_{q,q^{\prime}}(p_{i}) - M_{i}(q) \, S_{q,q^{\prime}}(p_{j}) \nn \\
                   & & + \sum_{k=1}  \binom{q^{\prime}}{k} \big(S_{q,k}(p_{i}) \, S_{q,q^{\prime}-k}(p_{j}) 
                       + S_{q,q^{\prime}-k}(p_{i}) \, S_{q,k}(p_{j})\big) \Biggr).
\label{2p_psd_addt}
\eea
where $M_{i}(q) = 1 - \sum_{i} p_{i}^{q}$ is the mixing between the various states and occurs due to the 
fractal nature of the phase space. 
The Shannon additivity relations corresponding to the various limiting cases of our two parameter entropy
(\ref{2p_entr}) are listed below for the sake of completeness. \\
{\it Special cases corresponding to the various one parameter entropies: } \\
{\it (i)} In the $q\rightarrow 1$ limit the generalized Shannon additivity corresponding to the
fractional entropy (\ref{entr_fractional}) is obtained 
\bea
S_{q^\prime}(p_{ij}) &=& \frac{1}{2} \bigg(\sum_{i}s_{q^\prime}(p_{i})
                           + \sum_{i} p_{i} \sum_{j} s_{q^\prime} (p(j|i)) \nn \\
                       & & + \sum_{i} \sum_{j}\sum_{k=1} \binom{q^{\prime}}{k}
                           \Big(s_{k}(p_{i})s_{q^\prime-k} (p(j|i))    
                           + s_{q^\prime-k}(p_{i})s_{k} (p(j|i))\Big)\bigg).
\label{frac_shn_addt} 
\eea
When the joint probability of system $p_{ij}$ obeys the relation $p_{ij} = p_{i} p_{j}$, we recover 
the pseudoadditivtiy relation proved in [\cite{MR98}].  \\
{\it (ii)} The generalized Shannon additivity corresponding to the fractal entropy (\ref{entr_fractal}) is recovered
in the $q^{\prime} \rightarrow 1$ limit
\beq
S_{q}(p_{ij})= \sum_{i} p_{i}^{q} \sum_{j} s_{q} (p(j|i)) + \sum_{i} s_{q}(p_{i}) \sum_{j} (p(j|i))^{q},
\label{fractal_shn_addt} 
\eeq
and the pseudoadditivity relation corrresponding to this entropy given in [\cite{FS04}] can be obtained under the 
condition $p_{ij} = p_{i} p_{j}$. \\
{\it (iii)} The generalized Shannon additivity relation corresponding to the one parameter entropy 
(\ref{Ub_Fr_entr}) in the $q^{\prime} \rightarrow q$ limit has the form
\bea
S_{q}(p_{ij}) &=& \frac{1}{2} \bigg(\sum_{i} s_{q}(p_{i}) \sum_{j} (p(j|i))^q 
                  + \sum_{i}\sum_{j}  \sum_{k=1} \binom{q}{k}
                  \Big(p_{i}^k (p(j|i))^{q-k} s_{q-k}(p_{i})s_{k} (p(j|i))  \nn \\
              & & + p_{i}^{q-k} (p(j|i))^{k} s_{k}(p_{i})s_{q-k} (p(j|i)) \Big)
                  + \sum_{i} p_{i}^q \sum_{j} s_{q}(p(j|i))\bigg).
\label{1p_shn_addt}
\eea
Imposing the condition $p_{ij}=p_{i} p_{j}$ we can get the following pseduoadditivity relation 
\bea
S_{q}(p_{ij}) &=& \frac{1}{2} \bigg(S_{q}(p_{i})+S_{q}(p_{j}) - M_{j}(q) S_{q}(p_{i}) - M_{i}(q) S_{q}(p_{j}) \nn \\
           & & + \sum_{k=1} \binom{q}{k} \bigg(S_{q-k}(p_{i}) S_{k}(p_{j}) (1-M_{i}(k))(1-M_{j}(q-k))  \nn \\
           & & + S_{k}(p_{i}) S_{q-k}(p_{j})(1-M_{i}(q-k))(1-M_{j}(k)) \bigg)\bigg).
\label{1p_pseud_addt}
\eea
\subsection{Uniqueness of the two parameter entropy}
In this subsection we prove the uniqueness of the two parameter entropy which obeys the modified form of the 
Shannon additivity given in (\ref{Gen_shn_addt}).  It can be noticed that (\ref{Gen_shn_addt}) is a symmetrized
combination of the following two equations
\bea
\sum_{i,j} s_{q,q^\prime}(p_{ij}) &=& \sum_{i,j} s_{q,q^\prime}(p_{i}) \big(p(j|i)\big)^{q}  
                           + \sum_{k=1}  \binom{q^{\prime}}{k}  \sum_{i,j} s_{q,q^\prime-k}(p_{i}) 
                            s_{q,k}\big(p(j|i)\big)
\label{bino_exp_11} \\
\sum_{i,j} s_{q,q^\prime}(p_{ij}) &=& \sum_{i,j} p_{i}^q  s_{q,q^\prime}(p(j|i)) 
                           + \sum_{k=1} \binom{q^{\prime}}{k}  \sum_{i,j} s_{q,k}(p_{i}) 
                            s_{q,q^\prime-k}\big(p(j|i)\big).
\label{bino_exp_21}
\eea
We come to this conclusion, since in (\ref{Gen_shn_addt}) there is a $p_{i} \leftrightarrow p(j|i)$ symmetry between 
the first and the second term and also between the two terms within the $k$ summation. Since the {\it lhs} in (\ref{bino_exp_11})
and (\ref{bino_exp_21}) are equal, the {\it rhs} of these equations should also have matching individual terms.   
This implies that $s_{q,0}(p_{i}) = p_{i}^{q}$ and also that the entropic function $s_{q,q^{\prime}}$ can be separated 
in the form of $\mathsf{s}_{q} \, \mathfrak{s}_{q^{\prime}}$. Since we already know that 
$\mathsf{s}_{q}(p_{i}) = s_{q,0}(p_{i}) = p_{i}^{q}$ what remains is to find the functional form of 
$\mathfrak{s}_{q^{\prime}}(p_{i})$.  Using the separable form of the entropy and the structure of $\mathsf{s}_{q}$ 
in (\ref{bino_exp_11}) we arrive at
\beq
\sum_{i,j}  p_{ij}^{q} \, \mathfrak{s}_{q^{\prime}}(p_{ij}) = \sum_{i,j} p_{ij}^{q} \, \mathfrak{s}_{q^{\prime}}(p_{i})
                                                              + \sum_{k=1} \binom{q^{\prime}}{k} \sum_{i,j} \, p_{ij}^{q} \,
                                                              \mathfrak{s}_{q^{\prime}-k}(p_{i}) \, \mathfrak{s}_{k}\big(p(j|i)\big).
\label{unq_sep_entr}
\eeq
This can be rewritten in the following form
\beq
\sum_{i,j}  p_{ij}^{q} \, \mathfrak{s}_{q^{\prime}}(p_{ij}) = \sum_{k=0} \binom{q^{\prime}}{k} \sum_{i,j} \, p_{ij}^{q} \,
                                                              \mathfrak{s}_{q^{\prime}-k}(p_{i}) \, \mathfrak{s}_{k}\big(p(j|i)\big).
\label{unq_bino_exp}
\eeq
Comparing the coefficients of $p_{ij}^{q}$ we get
\beq
\mathfrak{s}_{q^{\prime}}(p_{ij}) = \left(\mathfrak{s}(p_{i}) + \mathfrak{s}\big(p(j|i)\big)\right)^{q^{\prime}}.
\label{unq_ln_form}
\eeq
The only function which satisfies the form shown above is the logarithm.  The entropy is a positive function, whereas the 
probabilities can take only the values $0 \leq p_{i} \leq 1$, so this leads to the conclusion that $\mathfrak{s}_{q^{\prime}}(p_{i})
=  (\ln (1/p_{i}))^{q^{\prime}}$. Combining the two parts of the entropy $\mathsf{s}_{q}(p_{i})$ and $\mathfrak{s}_{q^{\prime}}(p_{i})$
we get (\ref{2p_entr}) the form of the entropy. 

Finally for the sake of completeness we define the conditional entropy of a pair of discrete random 
variables $(X,Y)$ with a joint distribution $p(x,y)$ as
\beq
S_{q,q^{\prime}} (X \lvert Y) = \sum_{x,y} p(x,y)^{q} (-\ln p(x|y))^{q^{\prime}},  
\eeq
where $p(x|y)$ denotes the conditional probability.
%
%
%
%
\setcounter{equation}{0}
\section{Generalized Divergence Measures}
\label{measures}
Divergence measures play an important role in Information theoretic analysis of any entropy, since
the probability distribution of a random variable cannot always be found exactly and also due to the 
reason that sometimes it is necessary to find the difference between two distributions. For the 
Shannon entropy several such measures like the Kullback Liebler relative entropy, Jensen Shannon divergence,
etc., have been introduced and investigated in detail. In this section we define these measures for the
two parameter entropy proposed in the previous section.  
\subsection{Relative entropy}
If $\mathsf{P}= \{ {\mathsf{p}}_{1}, \ldots , {\mathsf{p}}_{n} \}$ and 
$\mathfrak{P} = \{ {\mathfrak{p}}_{1}, \ldots , {\mathfrak{p}}_{n} \}$ be any two probability distributions, 
the two parameter relative entropy corresponding to these distributions is defined as
\beq
D_{q,q^{\prime}}(\mathsf{P} \| \mathfrak{P}) = \sum_{i} \mathsf{p}_{i}^{q} 
           \left(\ln \frac{\mathsf{p}_{i}}{\mathfrak{p_{i}}}\right)^{q^{\prime}},
\label{rel_entr_2p}
\eeq
where $q$ and $q^{\prime}$ are the generalizing parameters.  In the limit $q \rightarrow 1$ we recover the
fractional relative entropy and in the limit $q^{\prime} \rightarrow 1$ the fractal relative entropy can be obtained.
In the case where $q^{\prime} = q$ we obtain the Kullback relative entropy corresponding to the one parameter 
entropy described in (\ref{Ub_Fr_entr}). When both the parameters are set to unity we get back the Kullback relative 
entropy. 

\vspace{0.2cm}

\noindent Below we list the properties of the two parameter entropy and prove them. \\
{\it (i) Nonnegativity:} $D_{q,q^{\prime}}(\mathsf{P}  \| \mathfrak{P}) \geq 0$.  \\
{\it (ii) Continuity:}  $D_{q,q^{\prime}}(\mathsf{P} \| \mathfrak{P})$ is a continuous 
function for the 2n variables.  \\
{\it (iii)Symmetry:}
The relative entropy is symmetric under the simultaneous exchange of a pair of variables in the distributions
$\mathsf{P}$ and $\mathfrak{P}$ 
\bea
& & D_{q,q^{\prime}}({\mathsf{p}}_{1}, \ldots, {\mathsf{p}}_{j}, \ldots,{\mathsf{p}}_{k}, \ldots,{\mathsf{p}}_{n} 
\| {\mathfrak{p}}_{1}, \ldots, {\mathfrak{p}}_{j}, \ldots,{\mathfrak{p}}_{k}, \ldots,{\mathfrak{p}}_{n}) \nn \\
& & = 
D_{q,q^{\prime}}({\mathsf{p}}_{1}, \ldots, {\mathsf{p}}_{k}, \ldots,{\mathsf{p}}_{j}, \ldots,{\mathsf{p}}_{n}
\| {\mathfrak{p}}_{1}, \ldots, {\mathfrak{p}}_{k}, \ldots,{\mathfrak{p}}_{j}, \ldots,{\mathfrak{p}}_{n}).
\label{relentr_symm}
\eea
{\it (iv) Possibility of Extension:}
\beq
D_{q,q^{\prime}}({\mathsf{p}}_{1}, \ldots, {\mathsf{p}}_{n},0 \| {\mathfrak{p}}_{1}, \ldots, {\mathfrak{p}}_{n},0) 
= D_{q,q^{\prime}}({\mathsf{p}}_{1}, \ldots, {\mathsf{p}}_{n} \| {\mathfrak{p}}_{1}, \ldots, {\mathfrak{p}}_{n}).
\label{rel_entr_ext}
\eeq
{\it (v) Pseudoadditivity:}
\bea
D_{q,q^{\prime}}({\mathsf{P}}^{(1)} \times {\mathsf{P}}^{(2)} \| {\mathfrak{P}}^{(1)} \times {\mathfrak{P}}^{(2)}) &=& 
\frac{1}{2} \bigg(D_{q,q^{\prime}}({\mathsf{P}}^{(1)} \| {\mathfrak{P}}^{(1)}) 
- M_{j}(q) D_{q,q^{\prime}}({\mathsf{P}}^{(1)} \| {\mathfrak{P}}^{(1)})
\\ \nn
& & + D_{q,q^{\prime}}({\mathsf{P}}^{(2)} \| {\mathfrak{P}}^{(2)})
- M_{i}(q) D_{q,q^{\prime}}({\mathsf{P}}^{(2)} \| {\mathfrak{P}}^{(2)}) \\
& & + \sum_{k} \binom{q^{\prime}}{k} \Big(D_{q,k}({\mathsf{P}}^{(1)} \| {\mathfrak{P}}^{(1)}) \,
    D_{q,q^{\prime}-k}({\mathsf{P}}^{(2)} \| {\mathfrak{P}}^{(2)}) \nn \\
& & + D_{q,q^{\prime}-k}({\mathsf{P}}^{(1)} \| {\mathfrak{P}}^{(1)}) \,
      D_{q,k} ({\mathsf{P}}^{(2)} \| {\mathfrak{P}}^{(2)})\Big) \bigg). \nn
\eea
where
${\mathsf{P}}^{(1)} \times {\mathsf{P}}^{(2)} = \left\{a_{i}^{(1)} a_{i}^{(2)} \lvert a_{i}^{(j)} \in {\mathsf{P}}^{(j)}, j=1,2 \right \} $ and \\
${\mathfrak{P}}^{(1)} \times {\mathfrak{P}}^{(2)} = \left\{b_{i}^{(1)} b_{i}^{(2)} \lvert b_{i}^{(j)} \in {\mathfrak{P}}^{(j)}, j=1,2 \right \}$. \\
{\it (vi) Joint $q$-convexity:} 
\beq
D_{q,q^{\prime}}(\lambda {\mathsf{P}}^{(1)} + (1 - \lambda) {\mathsf{P}}^{(2)} 
\| \lambda {\mathfrak{P}}^{(1)} + (1 - \lambda) {\mathfrak{P}}^{(2)}) \leq 
\lambda^{q} D_{q,q^{\prime}} ({\mathsf{P}}^{(1)} \| {\mathfrak{P}}^{(1)}) + 
(1 - \lambda)^{q} D_{q,q^{\prime}} ({\mathsf{P}}^{(2)} \| {\mathfrak{P}}^{(2)}).
\label{q_convex} 
\eeq

\noindent {\it Proof:} \\
The convexity of the relative entropy function (\ref{rel_entr_2p}) proves the first axiom. Axioms {\it (ii)}, 
{\it (iii)} and {\it (iv)} can be trivially proved. The expression for pseudoadditivity in {\it (v)} follows 
from direct calculation.  The relative entropy satisfies the joint $q$-convexity stated in axiom {\it (vi)} 
proposed in Reference [\cite{AF09}].  To prove the joint $q$-convexity we use the generalized form of the 
log-sum inequality 
\beq
\sum_{i=1}^{n} \alpha_{i}^{q} \left( \ln \frac{\alpha_{i}}{\beta_{i}} \right)^{q^{\prime}} \geq
\left(\sum_{i=1}^{n} \alpha_{i}\right)^{q} \, \left(\ln \frac{{\displaystyle \sum _{i}^{n}} \alpha_{i}}
{{\displaystyle \sum_{i=1}^{n}} \beta_{i}} \right)^{q^{\prime}}.
\label{q_log_sum}
\eeq
This inequality can be obtained from $q$-generalization of Jensen inequality proposed in [\cite{AF09}].

Since the relative entropy (\ref{rel_entr_2p}) is not symmetric in $\mathsf{P}$ and $\mathfrak{P}$, we define
the following symmetric measure 
\beq
{\mathsf{J}}_{q,q^{\prime}}(\mathsf{P}, \mathfrak{P}) = \frac{1}{2} \, \left(D_{q,q^{\prime}} (\mathsf{P} \| \mathfrak{P}) + 
                              D_{q,q^{\prime}}(\mathfrak{P} \| \mathsf{P})\right),
\label{Gen_JKLd}
\eeq
which shares most of the properties of the relative entropy.  It can be noticed that the Kullback Liebler relative 
entropy (\ref{rel_entr_2p}) is undefined if the distribution ${\mathfrak{P}} = 0$ and ${\mathsf{P}} \neq 0$.  Similarly
the symmetric form of the relative entropy (\ref{Gen_JKLd}) is undefined if any of the distribution vanishes.  This 
implies that the distribution ${\mathfrak{P}}$ has to be continuous with respect to the distribution ${\mathsf{P}}$ 
for the measure (\ref{rel_entr_2p}) to be defined. For the case of the symmetric measure the distributions 
${\mathfrak{P}}$ and ${\mathsf{P}}$ have to continuous with respect to each other.  

To overcome this a modified form of the relative entropy is defined between the distribution $\mathsf{P}$ and a 
distribution which is a symmetric sum of both ${\mathsf{P}}$ and ${\mathfrak{P}}$. The mathematical expression 
corresponding to the modified relative entropy is
\beq
{\mathcal{D}}_{q,q^{\prime}} ({\mathsf{P}}, {\mathfrak{P}}) 
= D_{q,q^{\prime}} \left(\mathsf{P} \| (\mathsf{P} + \mathfrak{P})/2 \right)
= \sum_{i} \mathsf{p}_{i}^{q} 
           \left(\ln \frac{\mathsf{p}_{i}}{(\mathsf{p}_{i}+ \mathfrak{p}_{i})/2}\right)^{q^{\prime}}.
\label{symsm_KL_relentr}
\eeq
The alternative form of relative entropy proposed in the above equation is defined even when $\mathfrak{P}$ 
is not absolutely continuous with respect to the distribution $\mathsf{P}$.  Though it satisfies all the 
properties of the Kullback Liebler relative entropy it is not a symmetric measure.  So, a symmetric measure
based on (\ref{symsm_KL_relentr}) is defined as follows: 
\beq
{\mathcal{J}}_{q,q^{\prime}}(\mathsf{P}, \mathfrak{P}) = \frac{1}{2} \, \left({\mathcal{D}}_{q,q^{\prime}} (\mathsf{P} \| \mathfrak{P}) + 
                              {\mathcal{D}}_{q,q^{\prime}}(\mathfrak{P} \| \mathsf{P})\right).
\label{Gen_JKLd_mod} 
\eeq
which is a generalization of the Jensen Shannon divergence measure [\cite{JL91}] corresponding to the two parameter entropy 
(\ref{2p_entr}) defined in the previous section. 

The symmetric measure based on the relative entropy (\ref{Gen_JKLd}) and the generalized Jensen Shannon divergence measure 
(\ref{Gen_JKLd_mod}) satisfies the following properties
\bea
J_{q,q^{\prime}}(\mathsf{P}, \mathfrak{P}) &\geq& 0, \nn \\
J_{q,q^{\prime}}(\mathsf{P}, \mathfrak{P}) &=& J_{q,q^{\prime}}(\mathfrak{P},\mathsf{P}), \nn \\
J_{q,q^{\prime}}(\mathsf{P}, \mathfrak{P}) &=& 0  \Leftrightarrow  {\mathsf{P}} = {\mathfrak{P}},
\label{JSd_prop}
\eea
where $J_{q,q^{\prime}}$ can be either ${\mathsf{J}}_{q,q^{\prime}}$ or ${\mathcal{J}}_{q,q^{\prime}}$. The symmetric
form of the relative entropy and the Jensen Shannon divergence measure are related via the expression
\beq
{\mathcal{J}}_{q,q^{\prime}} \leq \frac{1}{2^{q^{\prime}}} \, {\mathsf{J}}_{q,q^{\prime}},
\label{JSd_modRE_rel}
\eeq
which clearly shows that the upper bound to the Jensen Shannon divergence is given by the symmetric form of the 
relative entropy.  A similar relationship also exists between the (\ref{symsm_KL_relentr}) and (\ref{rel_entr_2p}) 
in which the relative entropy defines the upper bound of the modified relative entropy. 
\subsection{Fisher information measure}
The Fisher information measure for a continuous random variable $X$ with probability distribution
$p(x)$ is defined as:
\beq
I = \int p(x) \left(\frac{1}{p(x)} \frac{d p(x)}{d x} \right)^{2} dx
  = \left\langle \left(\frac{1}{p(x)} \frac{d p(x)}{d x} \right)^{2}  \right\rangle
  = \left\langle \left(\frac{d}{d x} \ln p(x) \right)^{2}  \right\rangle.
\label{BG_fim}
\eeq
In Reference [\cite{GV95}], the Fisher information was obtained from the Kullback Liebler relative 
entropy in the following manner: The relative entropy between a uniform probability distribution 
$p(x)$ and its shifted measure $p(x+\Delta)$ is constructed.  The integrand is then expanded as 
a Taylor series in the shift $\Delta$ upto second order from which the Fisher information measure 
is recognized. Analogously we proceed to derive the $q,q^{\prime}$-generalized Fisher information 
measure using the two parameter generalized relative entropy proposed in the previous subsection. 
The two parameter relative entropy between the measure $p(x)$ and its shifted measure $p(x + \Delta)$ is   
\beq
D_{q,q^{\prime}} \big(p(x)\| p(x+\Delta)\big) 
= \int (p(x))^{q} \bigg(-\ln \frac{p(x+\Delta)}{p(x)}\bigg)^{q^{\prime}} dx. 
\eeq
In the above equation, the function $\ln p(x+\Delta)$ is expanded upto second order in $\Delta$ and 
the resulting expression is written in terms of a binomial series as follows: 
\beq
D_{q,q^{\prime}} \big(p(x) \|  p(x+\Delta)\big) 
= \int \frac{\Delta^{q^{\prime}}}{(p(x))^{-q}}   \sum_{k=0}^{\infty} \binom{q^{\prime}}{k} \bigg(\frac{p^{\prime}(x)}{p(x)}\bigg)^{q^{\prime}-k}
\bigg[ \frac{\Delta}{2} \bigg(\bigg(\frac{p^{\prime}(x)}{p(x)}\bigg)^{2} - \frac{p^{\prime \prime}(x)}{p(x)}\bigg)\bigg]^{k} dx. 
\eeq
Considering the first two lower order terms in $\Delta$  
\bea
D_{q,q^{\prime}} \big(p(x) \|  p(x+\Delta)\big) &=& \int (p(x))^{q} \bigg( \binom{q^{\prime}}{0} 
                                                          \Delta^{q^{\prime}} \bigg(\frac{p^{\prime}(x)}{p(x)}\bigg)^{q^{\prime}}  \nn \\
                                                      & & + \binom{q^{\prime}}{1} \frac{\Delta^{q^{\prime}+1}}{2} 
                                                          \bigg(\bigg(\frac{p^{\prime}(x)}{p(x)}\bigg)^{q^{\prime}+1}
                                                          -  \frac{\big(p^{\prime}(x)\big)^{q^{\prime}-1} p^{\prime \prime}(x)}{p^{q^{\prime}}(x)}\bigg)\bigg)
							  dx, \
\eea
and in comparison with the method adopted in Ref. [\cite{GV95}], we obtain the two parameter generalization
of the Fisher information measure 
\beq
I_{q,q^{\prime}} = \int (p(x))^{q-q^{\prime}-1} \ \left(\frac{d p(x)}{d x}\right)^{q^{\prime}+1} \, dx.
\label{2p_FIM}
\eeq
Along the lines of eqn (\ref{BG_fim}) the above expression can be defined through a $q$-expectation value as 
\beq
I_{q,q^{\prime}} = \left\langle \left(\frac{1}{p(x)} \frac{d p(x)}{d x} \right)^{q^{\prime}+1} \right\rangle_{q}
= \left\langle \left(\frac{d}{d x} \ln  p(x) \right)^{q^{\prime}+1} \right\rangle_{q}.
\label{qexpec_2p_FIM}
\eeq
The Fisher information measure (\ref{qexpec_2p_FIM}) make use of the $q$-expectation and in the $q^{\prime} \rightarrow 1$
limit it reduces to the expression obtained in [\cite{FP97}].
\subsection{Relative Fisher information and Jensen Fisher divergence measure: }
The relative Fisher information measure between two probability distributions $p_{1}(x)$ and $p_{2}(x)$ is given by
\beq
{}_{R}{I}_{q,q^{\prime}} (p_{1},p_{2}) =  \left\langle \left(\frac{{\rm d}}{{\rm d} x} 
                                          \ln \frac{p_{1}(x)}{p_{2}(x)}\right)^{q^{\prime}+1} \right\rangle_{q}
                                       =  \int (p_{1}(x))^{q} \left(\frac{d}{dx} \ln \frac{p_{1}(x)}{p_{2}(x)}\right)^{q^{\prime}+1}
                                          \, dx. 
\label{rel_FI_def}
\eeq
The above relation is not symmetric and so we define 
\beq
{\mathcal{I}}_{q,q^{\prime}}(p_{1},p_{2}) = \frac{1}{2}({}_{R}{I}_{q,q^{\prime}} (p_{1},p_{2}) + {}_{R}{I}_{q,q^{\prime}} (p_{2},p_{1})),
\label{rel_GFd}
\eeq
which is a symmetric extension of (\ref{rel_FI_def}). The disadvantage with these measures is the following: Equation 
(\ref{rel_FI_def}) requires the distribution $p_{2}(x)$ to be a continuous function of $p_{1}(x)$ for a given $x$ and 
relation (\ref{rel_GFd}) requires the distributions $p_{1}(x)$ and $p_{2}(x)$ continuous with respect to each other. 

To surmount this the relative Fisher information is defined between the distribution $p_{1}(x)$ and a new distribution
between $(p_{1} + p_{2})/2$ and the expression for this modified form reads: 
\beq
{\mathsf{I}}_{q,q^{\prime}}(p_{1},p_{2}) = {}_{R}{I}_{q,q^{\prime}} \left(p_{1},\frac{p_{1}+p_{2}}{2}\right)
                                         = \int (p_{1}(x))^{q} 
                                           \left(\frac{d}{dx} \ln \frac{p_{1}(x)}{(p_{1}(x) + p_{2}(x))/2}\right)^{q^{\prime}+1}
                                             \, dx.
\label{rel_FI_def_mod}
\eeq
A two parameter generalization of the recently proposed [\cite{PS12}] Jensen Fisher divergence measure, can be 
constructed via a symmetric combination of the modified form of the relative Fisher information.  The 
form of the generalized Jensen Fisher divergence measure so constructed is  
\beq
{\mathfrak{I}}_{q,q^{\prime}}(p_{1},p_{2}) = \frac{({\mathsf{I}}_{q,q^{\prime}} (p_{1},p_{2}) + {\mathsf{I}}_{q,q^{\prime}} (p_{2},p_{1}))}{2}
= \frac{({}_{R}{I}_{q,q^{\prime}} (p_{1},\frac{p_{1}+p_{2}}{2}) + {}_{R}{I}_{q,q^{\prime}} (p_{2},\frac{p_{1}+p_{2}}{2}))}{2}.
\label{rel_GFd_mod}
\eeq
The Jensen Fisher measure (\ref{rel_GFd_mod}) is convex and symmetric and vanishes only when both the 
probabilities $p_{1}(x)$ and $p_{2}(x)$ are identical everywhere.  

Finally we discuss the relevant limiting cases:  All the information measures defined in this section 
reduce to the corresponding measures obtained through the use of the Shannon entropy in the 
$q,q^{\prime} \rightarrow 1$. In the $q \rightarrow 1$ limit the information
measures relevant to the Fractional entropy is recovered.  The $q^{\prime} \rightarrow 1$ limit leads 
to the information measures of the Fractal entropy.   
%
%
%
%

\setcounter{equation}{0}
\section{Thermodynamic properties}
\label{thermal}
The canonical probability distribution $p_{i}$ can be obtained by optimizing the entropy subject to the 
the norm constraint and the energy constraint.  Adopting a similar procedure for our two parameter entropy 
(\ref{2p_entr}) we construct the functional
\beq
L= \sum \Phi (p_{i} ; q,q^{\prime}) - \alpha\bigg(\sum_{i}p_{i}-1 \bigg) -\beta \bigg(\sum_{i}p_{i} \epsilon_{i}-E \bigg),
\quad
\Phi(p_{i} ; q,q^{\prime}) = p_{i}^{q} (- \ln p_{i})^{q^{\prime}}, 
\label{optm_entr}
\eeq
where $\alpha$ and $\beta$ are the Lagrange's multiplier and $\epsilon_{i}$ is the energy eigenvalue and $E$ is the internal 
energy. 
Employing the variational procedure we optimize the functional in (\ref{optm_entr}) with respect to the probability to get
\beq
\frac{\delta L}{\delta p_{i}} =  \Phi^{\prime} (p_{i};q,q^{\prime}) - (\alpha + \beta \epsilon_{i}),
\label{max_entr} 
\eeq
where $\Phi^{\prime} (p_{i};q,q^{\prime})$ is 
\beq
\Phi^{\prime} (p_{i};q,q^{\prime}) = q p_{i}^{q-1} (-\ln p_{i})^{q^{\prime}} - q^{\prime} p_{i}^{q-1}(-\ln p_{i})^{q^{\prime}-1}.
\label{deriv_entr_func}
\eeq
When the functional $L$ attains a maximum its variation wrt $p_{i}$ is zero and using this in (\ref{max_entr}) yields the 
inverse of the probability distribution
\beq
q p_{i}^{q-1} (-\ln p_{i})^{q^{\prime}} - q^{\prime} p_{i}^{q-1}(-\ln p_{i})^{q^{\prime}-1} = 
(\alpha + \beta \epsilon_{i}).
\label{opt_func_1}
\eeq
Inversion of the relation (\ref{opt_func_1}) to obtain the probability distribution is not analytically feasible, so 
we adopt a different method to derive the distribution.  Since we have already set $\delta L /\delta p_{i}$ to zero,
we can integrate (\ref{max_entr}) to get 
\beq
\Phi (p_{i};q,q^{\prime}) = (\alpha + \beta \epsilon_{i}) \, p_{i}.
\label{phi_alpha_beta}
\eeq
Substituting the entropic expression (\ref{2p_entr}) in (\ref{phi_alpha_beta}) and comparing this with the equation for 
Lambert's $W$-function $z=we^{w}$ we obtain the relation for the probability 
\beq
p_{i} = \left(\frac{W(z)}{z}\right)^{\frac{q^{\prime}}{1-q}}, \qquad \qquad 
z = \frac{1-q}{q^{\prime}} \ (\alpha + \beta \epsilon_{i})^{\frac{1}{q^{\prime}}}.
\eeq
The factor $W(z)$ is the Lambert's $W$-function also known as the product log function.  For real $z$ the function
contains two branches denoted by $W_{0}(z)$ and $W_{-1}(z)$.  The branch $W_{0}(z)$ satisfies the condition that 
$W(z)\geq -1$ and is generally known as the principal branch of the $W$-function. When $W(z)\leq -1$ we have the 
$W_{-1}(z)$ branch. The Lambert's $W$-function occurs naturally in both classical [\cite{JC03}] and quantum statistical 
mechanics [\cite{SR09},\cite{JT10}] as well as in nonequilibrium statistical mechanics [\cite{EL05}]. Very recently 
its connection to the field of generalized statistical mechanics has been established through the following 
works [\cite{FS04},\cite{SA08},\cite{CJ12}], which our current result emphasizes.

\vspace{0.2cm}

\noindent{\it Lesche stability:}\\
A stability criterion was proposed by Lesche [\cite{BL82},\cite{BL04}] to study the stabilities of R\'enyi and the Boltzmann Gibbs entropy.
The motivation for this criterion goes as follows: An infinitesimal change in the probabilities $p_{i}$ 
should produce an equally infinitesimal changes in an observable. If $p$ and $p^{\prime}$ be two probability
distributions, Lesche stability requires that $\forall \epsilon > 0$ we can find a $\delta > 0$ such that 
\beq
\sum_{j=1}^{n} \mid p_{i}-p_{j}^{\prime}\mid \leq \delta \Rightarrow 
\frac{\mid S_{q,q^{\prime}}(p^{\prime})- S_{q,q^{\prime}}(p)\mid}{S_{q,q^{\prime}}^{max}} < \epsilon.
\label{Lesche_stab}
\eeq
Using (\ref{Lesche_stab}) a simple condition was derived in [\cite{SA04}] for any generalized entropy maximized by a 
probability distribution. This condition which is widely used to check the Lesche stabilities of generalized
entropies reads:
\beq
\frac{\mid S_{q,q^{\prime}}(p^{\prime})- S_{q,q^{\prime}}(p)\mid}{S_{q,q^{\prime}}^{max}} < 
C \sum_{j=1}^{n} \mid p_{i}-p_{j}^{\prime}\mid,
\label{LS_Chck_cond} 
\eeq
where the constant $C$ is 
\beq
C= \frac{f^{-1}(0^{+})- f^{-1}(1^{-})}{f^{-1}(0^{+}) - \int_{0}^{1} f^{-1}(p)dp}.
\label{LS_const_condt}
\eeq
The function $f^{-1}(p)$ is the inverse probability distribution obtained in (\ref{opt_func_1}).
In order to compute the constant we integrate the inverse probability distribution
with respect to the probability 
\beq
\int_{0}^{1} f^{-1}(p)dp= \int_{0}^{1} \big(q p^{q-1}(-\ln p)^{q^{\prime}}- 
q^{\prime} p^{q-1}(-\ln p)^{q^{\prime}-1} \big)dp.
\label{pinv_int}
\eeq
The {\it rhs} of (\ref{pinv_int}) consists of two integrals in $p$ and by using the 
transformation $\ln p= -y$ they are obtained in terms of the gamma function.  A simple 
calculation helps us to see that these two integrals are in fact the same and  so 
$\int_{0}^{1} f^{-1}(p)dp =0$. Similarly it can also be noticed that the $f^{-1}(1^{-})=0$
due to occurence of the natural logarithm.  So we finally get the value of $C$ as
\beq
C= \frac{f^{-1}(0^{+})}{f^{-1}(0^{+})}=1,
\label{LSC_fv}
\eeq
which leads to the conclusion that for our case $\delta = \epsilon$ and so the criterion for 
Lesche stability is satisfied.

\vspace{0.2cm}

\noindent{\it Thermodynamic stability:} \\
The thermodynamic stability conditions of the Boltzmann-Gibbs entropy can be derived from the maximum
entropy principle and its corresponding additivity relation.  In References [\cite{TW04},\cite{AS05}] it has been shown that
concavity alone does not guarantee thermodynamic stability for the two parameter entropy (\ref{2p_entr}).
We derive the stability conditions for the two parameter entropy {\it \'{a} la} the method developed 
in [\cite{AS05}].  The pseudoadditive relation for the two parameter entropy reads:
\bea
S_{q,q^{\prime}}(A,B) &=& \frac{1}{2} \bigg(S_{q,0}(A) \, S_{q,q^{\prime}}(B) + S_{q,0}(B) \, S_{q,q^{\prime}}(A)  + \sum_{k=1} \binom{q^{\prime}}{k}
                     \Big(S_{q,k}(A) \, S_{q,q^{\prime}-k}(B) \nn \\
                & &  + S_{q,q^{\prime}-k}(A) \, S_{q,k}(B) \Big) \bigg).
\label{psd_addt}            
\eea
Considering an isolated system comprising of two identical subsystems of energy $U$ in equilibrium, the total entropy of the 
system would be $S_{q,q^{\prime}}(U,U)$.  Allowing for an exchange of energy $\Delta U$ from one subsystem to the other 
subsystem the total entropy changes as $S_{q,q^{\prime}} (U+\Delta U, U-\Delta U)$, whose pseudoadditive relation 
following (\ref{psd_addt}) is
\bea
S_{q,q^{\prime}} (U+\Delta U, U-\Delta U) &=& \frac{1}{2} \bigg(S_{q,0}(U+\Delta U) S_{q,q^{\prime}}(U-\Delta U) \nn \\
                                          & &  + S_{q,0}(U-\Delta U) S_{q,q^{\prime}}(U+\Delta U) \nn \\
                                          & & + \sum_{k=1} \binom{q^{\prime}}{k}
                                               \Big(S_{q,k}(U+\Delta U) \, S_{q,q^{\prime}-k}(U-\Delta U)  \nn \\
                                          & &  + S_{q,q^{\prime}-k}(U+\Delta U) \, S_{q,k}(U-\Delta U) \Big) \bigg).
\label{psd_addt_eexch}            
\eea
Similarly, the pseudoadditive relation corresponding to $S_{q,q^{\prime}}(U,U)$ obtained using (\ref{psd_addt}) is
\bea
S_{q,q^{\prime}}(U,U) &=& \frac{1}{2} \bigg(S_{q,0}(U) S_{q,q^{\prime}}(U) + S_{q,0}(U) S_{q,q^{\prime}}(U)  + \sum_{k=1} \binom{q^{\prime}}{k}
                     \Big(S_{q,k}(U) S_{q,q^{\prime}-k}(U) \nn \\
                & &  + S_{q,q^{\prime}-k}(U)  S_{q,k}(U) \Big) \bigg).
\label{psd_addt_entr} 
\eea
From the maximum entropy principle we know that 
\beq
S_{q,q^{\prime}}(U,U) \geq S_{q,q^{\prime}} (U+\Delta U, U-\Delta U).
\label{max_entr_prp} 
\eeq
Expanding (\ref{psd_addt_eexch}) upto second order in $\Delta U$ and using the maximum entropy principle (\ref{max_entr_prp}),
we get the following condition 
\bea
0 &\geq& S_{q,0}\frac{\partial^{2} S_{q,q^{\prime}}}{\partial U^{2}} - \frac{\partial S_{q,0}}{\partial U} \frac{\partial S_{q,q^{\prime}}}{\partial U} 
+ \frac{1}{2} \sum_{k=1} \binom{q^{\prime}}{k} \bigg(S_{q,k} \frac{\partial^{2} S_{q,q^{\prime}-k}}{\partial U^{2}} 
+ S_{q,q^{\prime}-k} \frac{\partial^{2} S_{q,q^{\prime}}}{\partial U^{2}} \nn \\
& & - 2 \, \frac{\partial S_{q,k}}{\partial U} \ \frac{\partial S_{q,q^{\prime}-k}}{\partial U}\bigg).
\label{max_ent_con}
\eea
The two parameter entropy can be connected to its first and second derivatives via recurrence relations which 
be substituted in Eqn. (\ref{max_ent_con}) to yield the simplified form
\beq
0 \geq \frac{\partial^{2} S_{q,q^{\prime}}}{\partial U^{2}} - (1-q) \big((2^{q^{\prime}} - q)  S_{q,q^{\prime}} 
+ q^{\prime} (2^{q^{\prime}} + 2q + 1) \ S_{q,q^{\prime}-1} + ({q^{\prime}}^{2}+q^{\prime} + 2^{q^{\prime}})  S_{q,q^{\prime}-2}\big).
\label{therm_stab_con}
\eeq
From (\ref{therm_stab_con}) we notice that in the {\it rhs}, the first term is negative in the region $q,q^{\prime} > 0$
due to the concavity conditions imposed on the entropy.  From the rest of the terms we notice that the stability conditions
will be respected when either $0 < q < 1$ and $q^{\prime} > \log_{2} q$ or $q > 1$ and $q^{\prime} < \log_{2} q$.  Under
the limiting conditions of $q, q^{\prime} \rightarrow 1$, we recover the concavity condition for the Boltzmann entropy which 
is also the thermodynamic stability condition for the Boltzmann Gibbs entropy.  

\vspace{0.2cm}

\noindent{\it Generic example in the microcanonical ensemble:}\\
An isolated system in thermodynamic equilibrium can be described via the microcanonical ensemble. 
In a microcanonical picture all the microstates are equally probable.
Under conditions of equiprobability i.e., $p_{i} = \frac{1}{W} \, \forall i \in (1,...,W)$ 
the two parameter entropy (\ref{2p_entr}) becomes
\beq
 S_{q,q^{\prime}} = k \, W^{1-q} (\ln W)^{q^{\prime}}, 
 \label{Ub_entr_eqp}
\eeq
where $W$ is the total number of microstates. In the limit $q \rightarrow 1$ the above expression 
(\ref{Ub_entr_eqp}) reduces to the microcanonical entropy derived from the fractal entropy (\ref{entr_fractal}). Similarly 
in the $q^{\prime} \rightarrow 1$ limit we can obtain the entropic expression corresponding to the 
fractional entropy (\ref{entr_fractional}).  When we set $q^{\prime}=q$ we can get the microcanonical entropy 
corresponding to the entropy in (\ref{Ub_Fr_entr}). For the entropy (\ref{Ub_entr_eqp}) the temperature is defined 
through the relation
\beq
\frac{1}{T}= \frac{\partial S}{\partial E} =  k \ W^{-q} \ (\ln W)^{q^{\prime}-1} \big(q^{\prime} + (1-q) \ln W + q^{\prime}\big)
                              \frac{\partial W}{\partial E}.
\label{temp_def_entr}
\eeq
The definition of temperature corresponding to a generic class of systems for which the density of states $W$ is related
to the energy via the expression $W= C E^{f}$ is found to be 
\beq
\frac{1}{T}= k f C^{1-q} E^{(1-q)f-1} (\ln CE^{f})^{q^{\prime}-1} \big(q^{\prime}+ (1-q) \ln CE^{f} \big).
\label{temp_gc}
\eeq
An analytic inversion of (\ref{temp_gc}) to obtain the energy as a function of temperature is not feasible. Though the 
above illustration is given only for the microcanonical ensemble a direct extension of this method to include other
kinds of adiabatic ensembles can be easily achieved. 

%
%
%
%
\setcounter{equation}{0}
\section{Complexity Measures}
\label{complexity}
Physical systems in which the behaviour of the total system cannot be constructed from the properties of the 
individual components is generally defined as complex systems.  Several measures were proposed to quantify 
complexity of physical systems [\cite{SL88},\cite{RL95},\cite{JS99}]. One such measure is the 
disequilibrium based statistical measure of complexity popularly referred to as LMC (L\'{o}pez-Ruiz, Mancini and Calbet)
complexity measure which was introduced in [\cite{RL95}].  This measure is based on the logic that there are two
extreme situations in which we can find the simple systems, one is the perfect crystal in which the constituent atoms
are symmetrically arranged and the other limit is the completely disordered system which is best characterized by an 
ideal gas in which the system can be found in any of the accessible states with the same probability. The available 
information is very little in the case of a perfect crystal and is maximum for the ideal gas. The amount of information
in the system can be found using the Boltzmann Gibbs entropy $(S)$.  A new quantity called the disequilibrium $(\mathsf{D})$
was proposed which is the distance from the equiprobable distribution and is maximum in the case of the crystal and 
zero for an ideal gas.  The product of these quantities was defined as the measure of complexity. This measures 
vanishes for both perfect crystal and the ideal gas.

For a system consisting of $N$ accessible states with a set of probabilities $\{p_{1},p_{2},...,p_{N}\}$, 
obeying the normalization condition $\sum_{i=1}^{N} p_{i} = 1$ the complexity measure reads:
\beq
C = S  \ {\mathsf{D}}, \qquad  \qquad {\mathsf{D}} = \sum_{i=1}^{N}\left(p_{i} - \frac{1}{N}\right)^{2}.
\label{stat_compl}
\eeq 
The LMC measure of complexity was found to be a nonextensive quantity.  A generalized measure of complexity 
was proposed in [\cite{TY04}] based on Tsallis entropy with a view to absorb the nonadditive features of the entropy.
Similarly a statistical measure of complexity corresponding to the two parameter entropy (\ref{2p_entr}) is 
defined as follows:
\beq
C_{q,q^{\prime}} = S_{q,q^{\prime}}  \ {\mathsf{D}} 
                 \equiv \left(k \, \sum_{i}^{N} p_{i}^{q} \left(-\ln p_{i}\right)^{q^{\prime}}\right) 
                   \left(\sum_{i=1}^{N}\left(p_{i} - \frac{1}{N}\right)^{2}\right).
\label{q_stat_compl}
\eeq
In the limit $q,q^{\prime} \rightarrow 1$ we recover the LMC complexity measure proposed in [\cite{RL95}].
The LMC complexity measure corresponding to the fractal entropy and the fractional entropy are obtained in 
the $q^{\prime} \rightarrow 1$ and $q \rightarrow 1$ limits.  When we let $q^{\prime} = q$, the complexity
measure corresponding to the entropy (\ref{Ub_Fr_entr}) is recovered. 

As an example let us consider a two level system with probabilities $p$ and $(1-p)$.  The expression for the
entropy and the disequilibrium measure are as follows: 
\beq
S_{q,q^{\prime}}(p) = - \left(p \ln p + (1-p) \ln (1-p)\right), \qquad \qquad {\mathsf{D}}(p) = 2 (p-1/2)^{2}.
\label{entr_diseq_cal}
\eeq
The statistical complexity computed from these quantities  
\beq
C_{q,q^{\prime}}(p) = - 2 \left(p \ln p + (1-p) \ln (1-p)\right) (p-1/2)^{2},
\label{compl_cal}
\eeq
is plotted below for the sake of analysis.  
\begin{figure}[h!]
\begin{center}
\resizebox{77mm}{!}{\includegraphics{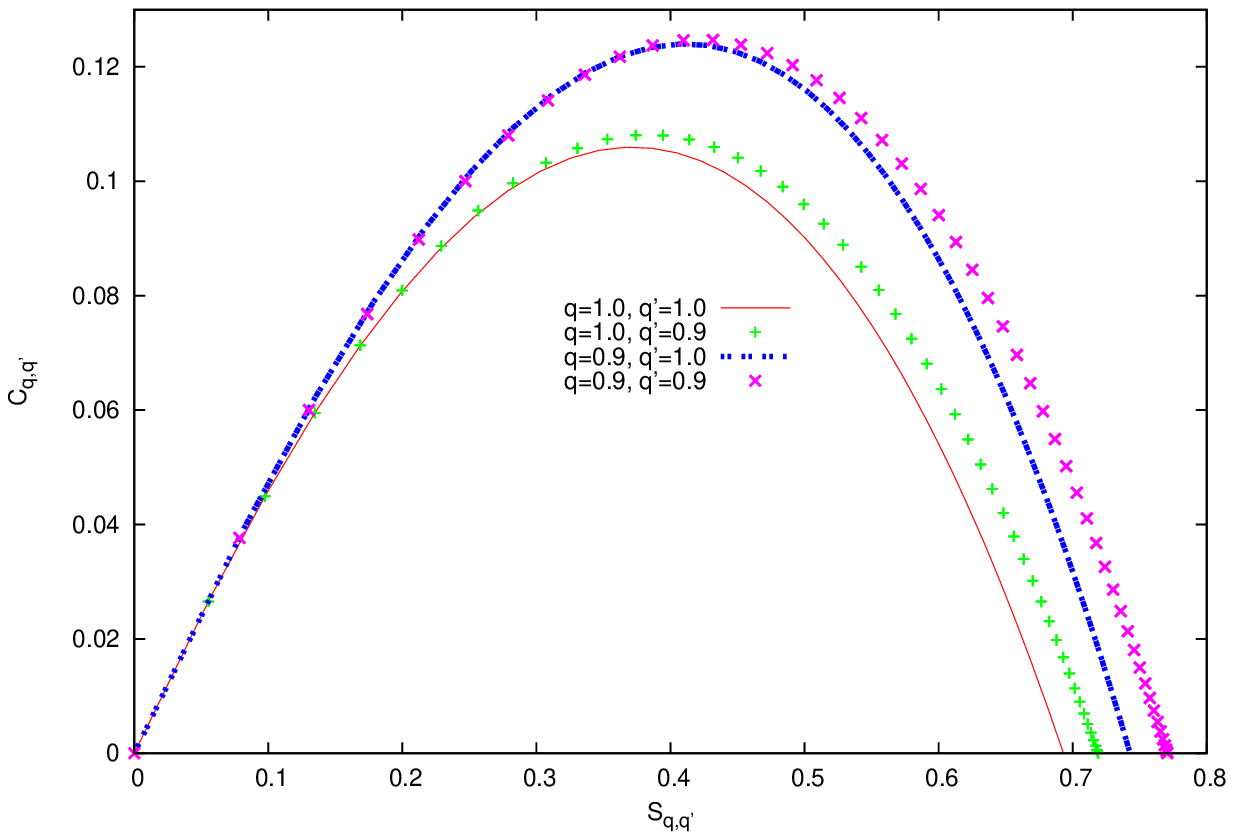}}
\resizebox{77mm}{!}{\includegraphics{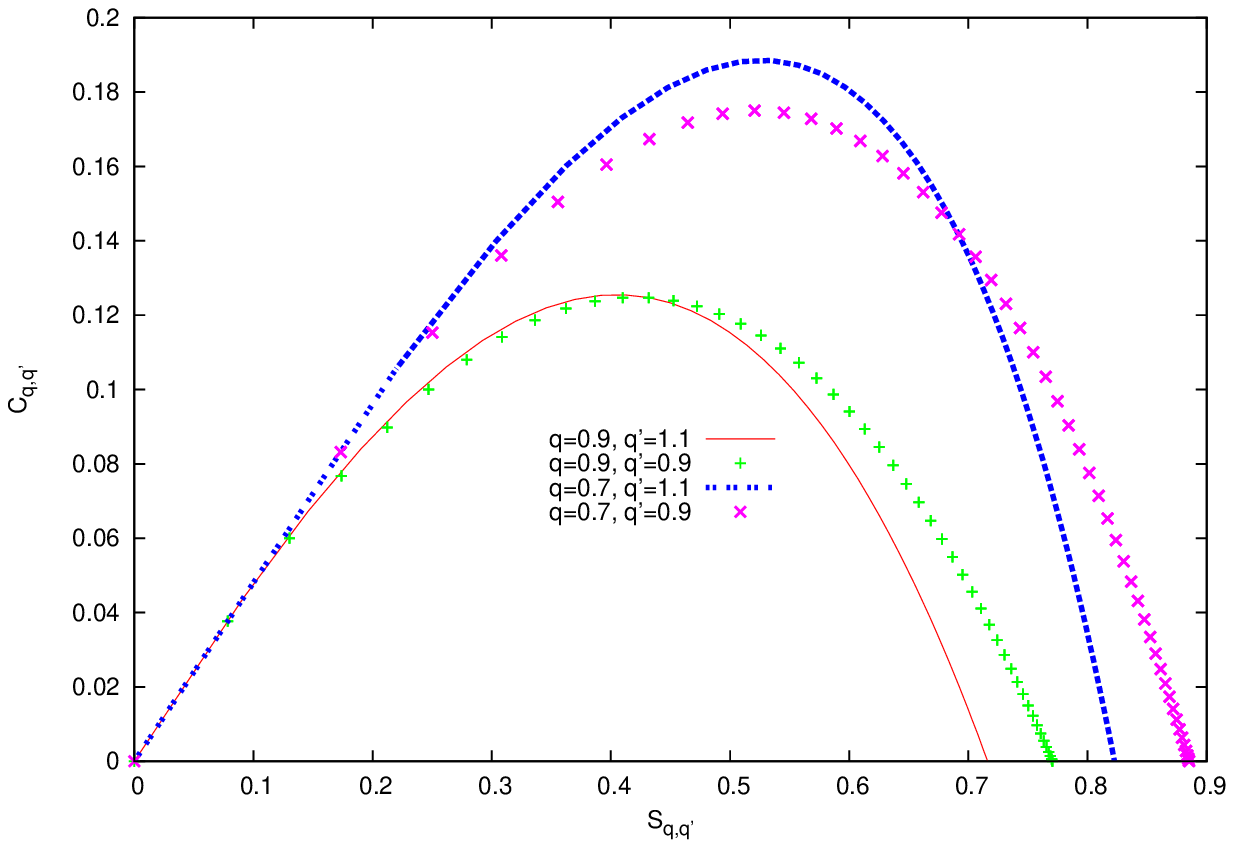}}
\end{center}
\caption{In the above graphs we plot the complexity measure $C_{q,q^{\prime}}$ with respect to the information 
$S_{q,q^{\prime}}$.  In the first plot we compare the LMC complexity measure for a two level system for the 
Boltzmann Gibbs entropy, the fractal entropy, the fractional entropy and the two parameter entropy (\ref{2p_entr}).
In the second graph we plot the LMC complexity measure for various values of the parameters $q$ and $q^{\prime}$.}
\end{figure}

From the plots we notice that the complexity measured using the Boltzmann Gibbs entropy, fractal entropy, 
fractional entropy and the two parameter entropy for the perfectly ordered state (crystal) goes to zero uniformly.
The largest complexity is achieved for the two parameter entropy followed by the fractal entropy, fractional entropy
and the Boltzmann Gibbs entropy. Also, the entropic value for which the maximum value of complexity is reached
differs for the different entropies.  The complexity zero corresponding to the disordered state occurs at various values 
of the entropy for the different entropies and the BG entropy reaches the zero first, followed by the fractional, the fractal
and the two parameter entropies.  For the two parameter entropy, the largest complexity is achieved more
quickly by varying the fractal index $q$ rather than the fractional index $q^{\prime}$. Also the zero complexity state
corresponding to the disordered system is attained more quickly when the fractional index $q^{\prime}$ is greater than one.

This measure can be extended to continuous probability distribution $p(x)$ obeying the normalization condition 
$\int_{-\infty}^{\infty} p(x) dx = 1$. For these distributions the summation over the states in the entropic definition 
is replaced by a integration over $x$. Similarly the disequilibrium measure is 
${\mathsf{D}} = \int_{-\infty}^{\infty}\left(p(x) - 1/N\right)^{2} dx$.  Since there is a continuum of states,
the number of states is very large and so the disequilibrium measure becomes ${\mathsf{D}} = \int_{-\infty}^{\infty}
\left(p(x)\right)^{2} dx$. The two parameter generalization of the LMC complexity measure for the continuum case 
is
\beq
C_{q,q^{\prime}} = S_{q,q^{\prime}}  \ {\mathsf{D}}
                 \equiv \left(k \, \int_{- \infty}^{\infty} (p(x))^{q} \left(-\ln p(x)\right)^{q^{\prime}} dx \right)
                 \left(\int_{- \infty}^{\infty} \left(p(x)\right)^{2} dx \right).
\label{com_cont_dist}
\eeq
For the purpose of illustration we consider the exponential distribution \\
$p(x) = \frac{1}{\lambda} \exp(- x/\lambda) (x, \lambda > 0)$ and calculate the corresponding two parameter
entropy and its disequilibrium measure 
\beq
S_{q,q^{\prime}} = \frac{\lambda}{q^{q^{\prime}+1}} \; \Gamma(q^{\prime}+1,q \ln \lambda), 
\qquad  \qquad
{\mathsf{D}} = \frac{1}{2 \lambda}.
\label{entr_deq_cont}
\eeq
The LMC complexity found by substituting (\ref{entr_deq_cont}) in (\ref{com_cont_dist}) reads: 
\beq
C_{q,q^{\prime}} = \frac{1}{q^{q^{\prime}+1}} \; \Gamma(q^{\prime}+1,q \ln \lambda),
\label{com_cont_fe} 
\eeq
where $\Gamma(a,b)$, is the incomplete gamma function in which the lower limit is replaced by a positive 
number.   In the limit $\lambda \rightarrow 1$, statistical complexity reduces to  \\
$C_{q,q^{\prime}} = \Gamma(q^{\prime}+1)/q^{q^{\prime}+1}$.
%
%
%
%
\setcounter{equation}{0}
\section{Conclusions}
\label{results}
A new two parameter entropy based on the natural logarithm and generalizing both the fractal entropy and the fractional
entropy is introduced. This encompasses an interesting limiting case, where the $N$-particle entropy can be expressed in 
terms of a sum of single particle biased Boltzmann entropies. The generalized form of the Shannon-Khinchin axioms are
proposed and verified for this new two parameter entropy. These axioms uniquely characterize our new entropy. The corresponding
Kullback Liebler relative entropy is proposed and its properties are investigated.  Utilizing the relative entropy a 
generalization of the Jensen-Shannon divergence is also achieved.  Exploiting the relative entropy between a probability
measure and its shift, we derive the generalized Fisher information. Also, we obtain generalized forms of the relative 
Fisher information and the Jensen-Fisher information. The Lesche stability, and the thermodynamic stability are verified
for our entropy and we also find that the canonical probability distribution which optimizes our entropy can be expressed
via the Lambert's $W$-function.  Finally we introduce a generalization of the LMC complexity measure making use of our 
two parameter entropy and apply it to measure the complexity of a two level system.  The results obtained indicate that
there is a change in the complexity value with the dominant contribution coming from the fractal index $q$.  Finally 
we also examine an exponential distribution as an example of a continuous probability distribution and compute the 
complexity measure.  

Though we have introduced a new two parameter entropy based on natural logarithm, we do not know the specific systems 
where this can be applied.  But from our investigations we assume that it will be of use in measuring complexity in fractal 
systems and systems which exhibit fractional dynamics in phase space. Towards this end investigating the complexity of 
probability distributions corresponding to the fractional diffusion equation [\cite{RM04}]  will be worth pursuing.  
%
%
%
%
%
\section*{Acknowledgements}
One of the authors CR would like to acknowledge the use of Library facilities at the Institute of Mathematical
Sciences (IMSc).  

%
%
%


\begin{thebibliography}{99}
%
\bibitem{AR61} A. R\'enyi, Proceedings of the $4^{th}$ Berkeley Symposium on Mathematics, Statistics and 
Probability, 547 (1960). 
%
\bibitem{DM75} D.P. Mittal, Metrika {\bf 22}, 35 (1975).
%
\bibitem{BS75} B.D. Sharma and I.J. Taneja, Metrika {\bf 22}, 205 (1975).
%
\bibitem{CT88} C. Tsallis, J. Stat. Phys. {\bf 52}, 479 (1988).
%
\bibitem{BC02} B.J.C. Cabral and C. Tsallis, Phys. Rev. {\bf E 66}, 0615101(R) (2002).
%
\bibitem{AP04} A. Pluchino, V. Latora and A. Rapisarda, Continuum Mech. Thermodyn. {\bf 16}, 245 (2004).
%
\bibitem{AM11} A.M. Mariz and C. Tsallis, {\it Long memory constitutes a unified mesoscopic mechanism
consistent with nonextensive statistical mechanics}, arXiv No: 1106.3100 [cond-mat.stat-mech]
%
\bibitem{QAW03} Q.A. Wang, Entropy {\bf 5}, 220 (2003). 
%
\bibitem{MR98} M.R. Ubriaco, Phys. Lett. {\bf A 373}, 2516 (2009).
%
\bibitem{MR09} M.R. Ubriaco, Phys. Lett. {\bf A 373}, 4017 (2009).
%
\bibitem{RL95} R. L\'{o}pez-Ruiz, H.L. Mancini and X. Calbet, Phys. Lett. {\bf A 209}, 321 (1995).
%
\bibitem{FS04} F. Shafee, {\it A New Nonextensive Entropy}, arXiv No: 0406044 [nlin.AO] (2004).
%
\bibitem{FS09} F. Shafee, {\it Generalized Entropy from Mixing: Thermodynamics, Mutual Information
and Symmetry Breaking}, arXiv No: 0906.2458 [cond-mat.stat-mech] (2009).
%
\bibitem{GK01} G. Kaniadakis, Physica {\bf A 296}, 405 (2001).
%
\bibitem{AL07} A. Lavagno, A.M. Scarfone and P. Narayanaswamy, J. Phys. {\bf A}:
Math. Theor {\bf 40}, 8635 (2007).
%
\bibitem{EJ65} E.T. Jaynes, Am. J. Phys. {\bf 33}, 391 (1965). 
%
\bibitem{AF09} A.F.T Martins, N.A. Smith, E.P. Xing, P.M.Q. Aguiar and M.A.T. Figueiredo,
               Journal of Machine Learning Research {\bf 10}, 935 (2009). 
%
\bibitem{JL91} J. Lin, IEEE Transactions on Information theory {\bf 37}, 145 (1991). 
%
\bibitem{GV95} G.V. Vstovsky, Phys. Rev. {\bf E51}, 975 (1995). 
%
\bibitem{PS12} P. S\`{a}nchez-Moreno, A. Zarzo and J.S. Dehesa, J. Phys. {\bf A}: Math. Theor.
               {\bf 45}, 125305 (2012).
%
\bibitem{JC03} J.M. Caillol, J. Phys. {\bf A}: Math. Theor. {\bf 36} 10431 (2003).
%
\bibitem{SR09} S.R. Valluri, M. Gil, D.J. Jeffrey and S. Basu, J. Math. Phys. {\bf 50}, 102103 (2009).
%
\bibitem{JT10} J. Tanguay, M. Gil, D.J. Jeffrey and S. Basu, J. Math. Phys. {\bf 51}, 123303 (2010). 
%
\bibitem{EL05} E. Lutz, Am. J. Phys. {\bf 73}, 968 (2005).  
%
\bibitem{SA08} S. Asgarani and B. Mirza, Physica {\bf A 387},  6277 (2008). 
%
\bibitem{CJ12} R. Chandrashekar and J. Segar {\it Adiabatic thermostatistics of the two parameter entropy 
and the role of Lambert's $W$-function in its applications} arXiv No: 1210.5499 [cond-mat.stat-mech]
%
\bibitem{FP97} F. Pennini and A. Plastino, Physica {\bf A 247}, 559 (1997).
%
\bibitem{BL82} B. Lesche, J. Stat. Phys. {\bf 27}, 419 (1982).
%
\bibitem{BL04} B. Lesche, Phys. Rev. {\bf E 70}, 017102 (2004).
%
\bibitem{SA04} S. Abe, G. Kaniadakis, A.M. Scarfone, J. Phys. {\bf A}: Math. Gen. {\bf 37}, 10513 (2004).
%
\bibitem{TW04} T. Wada, Physica {\bf A 340}, 126 (2004).
%
\bibitem{AS05} A.M. Scarfone and T. Wada, Phys. Rev. {\bf E  62}, 026123 (2005).  
%
\bibitem{SL88} S. Lloyd and H. Pagels, Ann. Phys. {\bf 188}, 186 (1988).
%
\bibitem{JS99} J. Shiner, M. Davison and P.T. Landsberg, Phys. Rev. {\bf E 59}, 1459 (1999).
%
\bibitem{TY04} T. Yamano, J. Math. Phys. {\bf 45}, 1974 (2004).
%
\bibitem{RM04} R. Metzler and J. Klafter, J. Phys. {\bf A}: Math. Theor. {\bf 37} R161 (2004).
\end{thebibliography}
\end{document}